\documentclass[aps,floats,twocolumn,epsf,prl,showpacs]{revtex4-1}
\usepackage{graphicx}

\begin{document}

\title{Divergent Precursors of the Mott-Hubbard Transition at the Two-Particle Level}
\author{T. Sch\"afer$^a$, G. Rohringer$^{a}$, O. Gunnarsson$^b$, S. Ciuchi$^{c}$,  G. Sangiovanni$^d$, A. Toschi$^a$}
\affiliation{$^a$Institute of Solid State Physics, Vienna University of Technology, 1040 Vienna, Austria}
\affiliation{$^b$ Max Planck Institute for Solid State Research, Stuttgart, Germany}
\affiliation{$^c$ Dipartimento di Scienze Fisiche e Chimiche, Universit\`a de L’ Aquila, Via Vetoio I-67010 Coppito-L’ Aquila and Istituto dei Sistemi Complessi, CNR, Italy}
\affiliation{$^d$ Institute of Physics and Astrophysics, University of W\"urzburg, W\"urzburg, Germany}

\date{ \today }

\begin{abstract}
Identifying  the fingerprints of the Mott-Hubbard
metal-insulator transition may be quite elusive  in correlated
metallic systems if the analysis is limited to the
single particle level.  However, our dynamical mean-field
calculations demonstrate that the situation changes completely if the
frequency dependence of the two-particle vertex functions is considered: 
The first non-perturbative precursors of the Mott physics are unambiguously 
identified well inside the metallic regime by the
divergence of the local Bethe-Salpeter equation in the charge
channel.  At low temperatures this occurs in the region where incoherent high-energy features emerge in the spectral function, while at high temperatures it is traceable up to the atomic-limit.
 \end{abstract}

\maketitle

\let\n=\nu \let\o =\omega \let\s=\sigma


\vskip 5mm

\noindent
{\sl Introduction}. Among all  fascinating phenomena characterizing the physics of correlated electronic systems, one of the most important is undoubtedly the
Mott-Hubbard metal-to-insulator transition (MIT)\cite{MH}. Here, the
onset of an insulating state is a direct consequence of the strong Coulomb
repulsion, rather than of the underlying electronic bandstructure. 
Mott MITs have been indeed identified in several correlated
materials\cite{ImadaREV}, especially in the class of transition
metal oxides and heavy fermions. The interest in the Mott MIT is not limited however to the transition ``per se'',  but it is also for
 the correlated (bad) metallic regime in its proximity.               
 In fact, this region of the phase-diagram often displays
a rich collection of intriguing or exotic phases, that many relate
also to the physics of the  high-temperature superconducting cuprates. 

An exact theoretical description of the Mott MIT represents a considerable
challenge due to its intrinsically non-perturbative
nature in terms of the electronic interaction. However, a
significant progress was achieved with the invention of the dynamical
mean field theory (DMFT)\cite{DMFTREV}.
By an accurate treatment of {\sl local} quantum correlations, DMFT
has allowed for the first non-perturbative analysis of the
Mott-Hubbard MIT in the  Hubbard model\cite{Hubbard}, and, 
in combination with ab-initio methods\cite{LDADMFTREV}, also for the
interpretation and the prediction of experimental spectroscopic
results for strongly correlated materials, such as, e.g., the paramagnetic phases of V$_2$O$_3$. Theoretically, 
several ``hallmarks'' of the onset of the Mott insulating phase can
be unambiguously identified in  DMFT: At the
one-particle level, a divergence of the local
electronic self-energy in the zero-frequency limit
is observed,  reflecting the opening of the Mott spectral gap, while, at the
two-particle level,  the local spin susceptibility
($\chi_s(\omega\! = \! 0)$) diverges at $T\! =\!0$, due to the onset of
long-living local magnetic moments in the Mott phase. 
 
\noindent
{\sl Description of the problem.} While the characterization of the MIT itself is quite clear, at least on a DMFT level,  the physics of the correlated metal regime in the vicinity of the MIT is far from being trivial and presents several
anomalies. We recall here: the
occurrence of kinks of purely electronic
origin\cite{kinks2007} in the angular resolved one-particle
spectral functions or in the electronic specific heat, the formation
of large instantaneous magnetic moments, screened by the metallic
dynamics\cite{Toschi2012}, the abrupt change of the out-of-equilibrium
behavior  after a quench of the electronic
interaction\cite{Eckstein2009}, and the changes in the energy-balance 
between the paramagnetic and the low-temperature (antiferromagnetically) ordered phase, which also affect the restricted optical sum-rules\cite{energybal}.
Partly motivated by these observations, many DMFT calculations have been aiming at general characterization of this regime, e.g., by studying
the phase-diagram of the half-filled Hubbard model.  However,  no trace of other
phase transitions has been found beyond the MIT itself and the (essentially) mean-field antiferromagnetically ordered phase, which is not of interest here.
Hence, one of the main outcomes of the preceding DMFT analyses, mostly
focusing on the evolution of one-particle spectral properties (and, to less extent, on susceptibilities\cite{Raas2009}), has been the definition of the ``borders'' of the so-called {\sl crossover} regions at
higher $T$ than those where the MIT can be observed. The shape
of these crossover regions has been analyzed in many different ways\cite{DMFTREV,Bullarev,bluemer_thesis,Ciuchi2006,Dobro2011}. We note here, that the (different)  criteria used for defining crossover regimes imply a certain degree of arbitrariness. Furthermore, the crossover region is located at much higher $T$s than those, where some of the abovementioned anomalies are observed.

In this paper, going {\sl beyond} the standard, typically one-particle, DMFT analyses, we present a completely unambiguous criterion to distinguish the ``weakly'' and the ``strongly''-correlated regions in the phase-diagram.  By studying the frequency structure of the two-particle local vertex functions of DMFT, we observe the divergence of the local
Bethe-Salpeter equation in the charge channel. This divergence defines
a regime  remarkably different, also in the shape, from the crossover region, where non-perturbative precursor effects of the MIT become active, even well 
inside the low-temperature metallic phase. 
The precise definition of such a regime represents a much
better playground for a general interpretation of the anomalous
physics emerging as a precursor of the MIT. Furthermore,  our analysis,
showing the occurrence of peculiar divergent features in some of the
two-particle local vertex functions of DMFT is  also expected to have a
significant impact on future calculations for strongly correlated
electron systems, because the two-particle local vertex functions 
represent a crucial ingredient for both (i)  the calculation of dynamical momentum-dependent susceptibilities in DMFT \cite{DMFTREV,Kunes2011,Park2011}, as well as (ii) the diagrammatic extensions\cite{DGA,DF} of the DMFT, aiming at the inclusion of non-local spatial correlations.

\noindent
{\sl DMFT results at the two particle-level.} We consider the Hubbard model on a square lattice  in the paramagnetic phase at half-filling, that is one of the most basic realizations  of the MIT in DMFT. The corresponding Hamiltonian is
\begin{equation}
H=-t\sum_{\langle ij\rangle \sigma }c_{i\sigma }^{\dagger }c_{j\sigma           
}+U\sum_{i}n_{i\uparrow }n_{i\downarrow},  \label{eq:Hubmod}
\end{equation}%
where $t$ is the hopping amplitude between nearest-neighbors, $U$ is
the local Coulomb interaction, and $c_{i\sigma }^{\dagger }$($c_{i\sigma }$) creates
(annihilates) an electron with spin $\sigma=\uparrow,\downarrow$ at site $i$; $n_{i\sigma          
}\!=\!c_{i\sigma }^{\dagger }c_{i\sigma }$. Hereafter, all energy
scales will be given in units of $D=4t=1$, 
i.e., of the half of the standard deviation of the non-interacting
DOS\cite{note_D}.

\begin{figure*}[t!]
        \centering
                \includegraphics[width=\textwidth,angle=0]{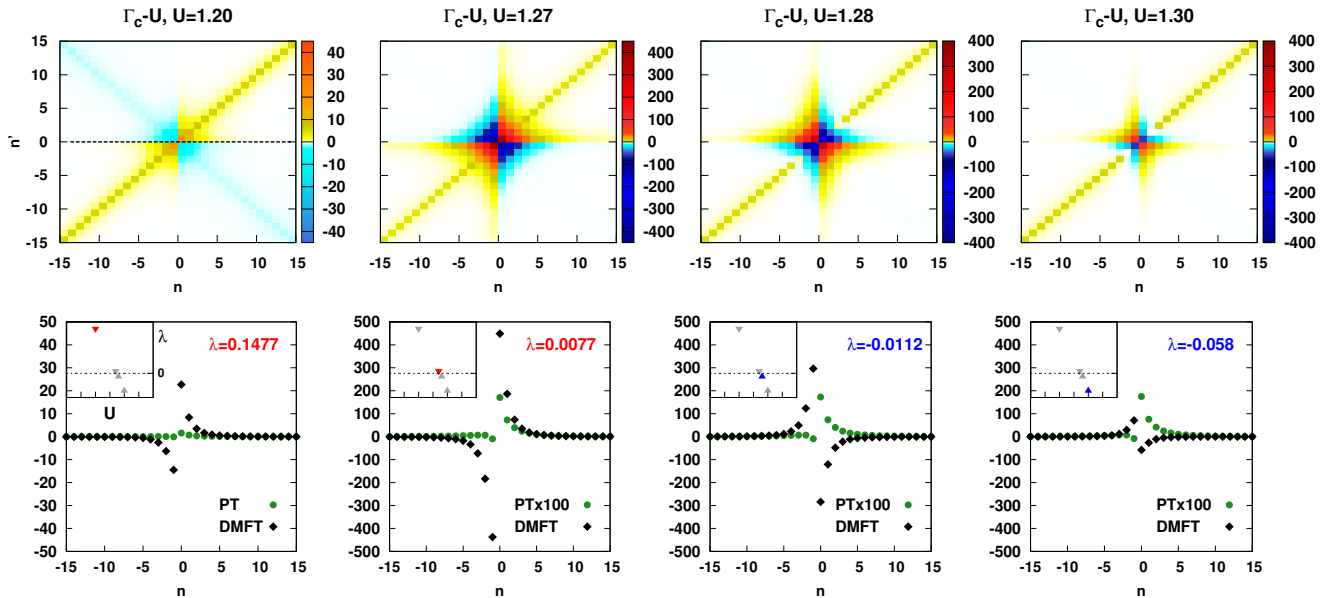}
        \caption{ \label{fig:1} (Color Online) {\sl Upper row}: Evolution of the frequency dependent two-particle
          vertex function, irreducible in the charge channel, ($\Gamma_c^{\nu \nu'}$)  for increasing
         $U$. The 
          data have been obtained by DMFT at zero external frequency
          ($\omega\!=\!0$) and fixed temperature ($T
          =0.1$); {\sl lower row}: linear snapshot of the same
          $\Gamma_c$ along the path marked by the dashed-line in the first panel of the upper row, i.e.,  as a function of
           $\nu=\frac{\pi}{\beta}(2n+1)$ for $n'=0$ ($\nu'=
           \frac{\pi}{\beta}$), compared to perturbation theory (PT) results. In the legends/insets the closest-to-zero eigenvalue ($\lambda$) of $\chi_{c}^{\nu,\nu'}/\chi_0^{\nu,\nu'}$ is reported for each $U$.}
\end{figure*}

Differently from previous studies, we will focus  on the
analysis of the two-particle local vertex functions computed with
DMFT. By using an Hirsch-Fye quantum Monte Carlo impurity solver\cite{DMFTREV}, whose accuracy has been also tested in selected cases with
exact-diagonalization DMFT calculations, we have first computed the 
generalized local susceptibility  $\chi^{\nu\nu'}(\omega)$. This is
defined, following the notation of Ref. \cite{Rohringer2012}, as:
\begin{eqnarray}
\chi^{\nu\nu'}_{\sigma\sigma'}(\omega) & = & \int d\tau_1 d\tau_2 d\tau_3 \,
e^{-i\nu\tau_1} e^{i(\nu+\omega)\tau_2} e^{- i(\nu'+\omega)\tau_3}  \\ \nonumber
    & \times & \left[ \langle T_\tau c^\dagger_\sigma(\tau_1) c_\sigma(\tau_2)
    c^\dagger_{\sigma'}(\tau_3) c_{\sigma'}(0)\rangle  \right. \\ \nonumber
    &    -  & \left.  \langle T_\tau c^\dagger_\sigma(\tau_1)
    c_\sigma(\tau_2) \rangle
     \langle T_\tau   c^\dagger_{\sigma'}(\tau_3) c_{\sigma'}(0)\rangle \right],
\label{eq:chi}
\end{eqnarray}
where $T_\tau$ is the (imaginary) time ordering operator, and
$\nu,\nu'$ and $\omega$ denote the two fermionic and the
bosonic Matsubara frequencies, respectively. Then, the Bethe-Salpeter
equation in the charge channel (defined as
$\chi^{\nu\nu'}_{c}(\omega)=\chi^{\nu\nu'}_{\uparrow\uparrow}(\omega)+
\chi^{\nu\nu'}_{\uparrow\downarrow}(\omega)$
has been considered for $\omega=0$, allowing to determine 
the corresponding irreducible vertex\cite{Rohringer2012}
\begin{equation}
\Gamma_{c}^{\nu\nu'} = [\chi_{c}^{\nu \nu'}(\omega=0)]^{-1}- [\chi_0^{\nu \nu'}(\omega=0)]^{-1},
\label{eq:Gamma}
\end{equation}
 where the last term is defined through the convolution of two DMFT
 Green's functions, as $\chi_0^{\nu \nu'}(\omega)= - T^{-1} G(\nu)
 G(\nu+\omega) \delta_{\nu \nu'}$. The vertex $\Gamma_{c}^{\nu\nu'}$ can be viewed as the two-particle counter-part of the electronic self-energy and, for a half-filled system, it is a purely real function. Our numerical results are reported in Fig. \ref{fig:1} for  four different values of the
electronic interaction $U$ at a fixed temperature of $T=0.1$. 
Starting to examine the first panel, corresponding to the
smallest value of $U=1.2$, one observes two main diagonal structures
in the Matsubara frequency space: These structures are easily 
interpretable  as originated by
reducible ladder-processes in the (transverse) particle-hole ($\nu \! = \! \nu'$) and in the particle-particle ($\nu \! = \!-\nu'$) channels respectively\cite{Rohringer2012}. Following
the behavior of the local spin susceptibility in the Mott phase, the main diagonal structure will diverge exactly at the MIT ($U_{\text{MIT}} \sim 3$) in the
$T=0$ limit.

\begin{figure}[t!]
          \vspace{2mm}
        \centering\includegraphics[width=0.45\textwidth,angle=0]{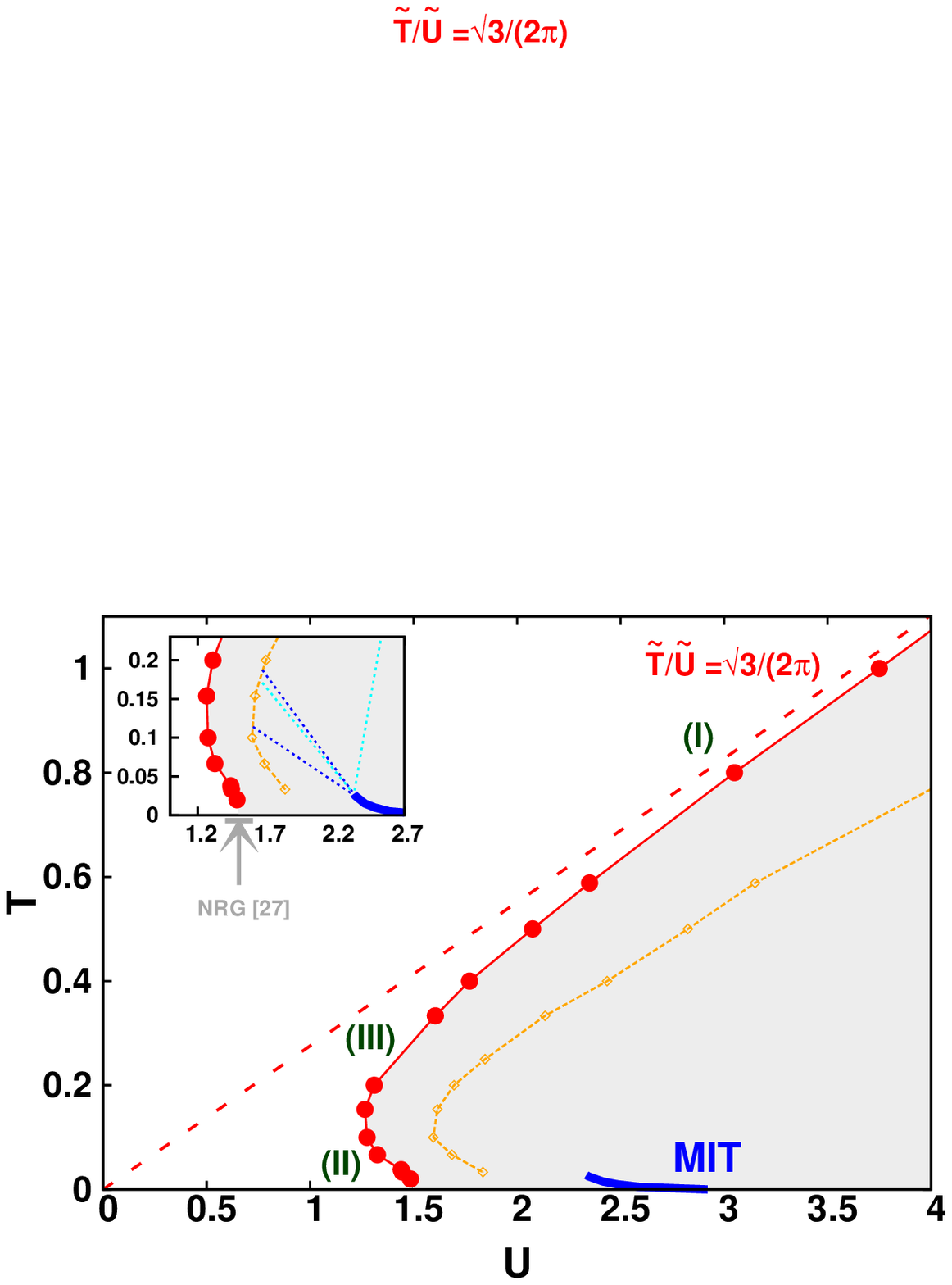}
        \caption{\label{fig:2} (Color Online)  Instability lines
          of the irreducible vertices in the charge ($\Gamma_c$
          red circles) and in the particle-particle 
          channels ($\Gamma_{pp}$ orange diamonds) reported in the DMFT
          phase diagram of the half-filled Hubbard model (the data of
        the MIT, blue solid line, 
      are taken from Ref. \cite{Bullarev,bluemer_thesis}). The red
      dashed line indicate the corresponding instability 
     condition ($\widetilde{T} = \frac{\sqrt{3}}{2\pi}\widetilde{U}$) estimated from the atomic limit. Inset: zoom on the low-$T$ region, where also different estimations (dashed light-blue\cite{Bullarev}, dashed blue\cite{bluemer_thesis}) of the crossover region are indicated.}
\end{figure}

In contrast to these standard properties of $\Gamma^{\nu\nu'}_{c}$,
visible in the first panel, the
analysis of the other three panels of Fig.\ \ref{fig:1} shows the
emergence of a low-frequency singular behavior of the vertex functions
for a value of $U$ much smaller than that of the MIT: Already at
$U=1.27$ (second panel), one observes a strong enhancement of the
 vertex function at the lowest Matsubara frequencies (note the change in the intensity-scale). This is visible as an emergent ``butterfly''-shaped  structure, where the intense red-blue color coding indicates alternating signs in the ($\nu$,$\nu'$) space. Remarkably, such a low-energy structure becomes 
predominant over the other ones along the diagonals.
That a true divergence takes place is suggested by the third panel
($U=1.28$),  where the intensity of  the``butterfly'' structure is equally
strong but the signs are now inverted as indicated by the colors. This is also shown more quantitatively by a selected cut of
$\Gamma_c^{\nu \nu'}$ in frequency space, reported in the corresponding
panels of the second row of Fig. 1. Note that the inversion of the
signs {\sl cannot} be captured by perturbation theory calculations (green
circles), marking the non-perturbative nature of the result.
The rigorous proof of the divergence is provided by the
evolution of the matrix $\chi_{c}^{\nu \nu'}$, which is positive definite at weak-coupling,
while one of its eigenvalues (see legends and insets in the bottom row of
Fig. \ref{fig:1}) becomes {\sl negative}  crossing
$0$.  Finally, by further increasing $U$, the low-energy
structure weakens, indicating that at fixed $T=0.1$ this vertex divergence is taking place only for a specific value of the Hubbard interaction, i.e., for
$\widetilde{U} \simeq 1.275$. 
This finding naturally leads to the crucial
question of the temperature dependence of the results: Does such
divergence occur for all temperatures, and if yes, is the temperature
dependence of $\widetilde{U}$ significant? As one can immediately
understand from Fig.\ref{fig:2}, the answer to both questions
is positive\cite{note_Tdep}: By repeating the analysis of Fig.\ref{fig:1} at
different temperatures, we could identify  
the loci ($\widetilde{T},\widetilde{U}$, red dots in Fig.\ref{fig:2}) in the phase-diagram, where the low-frequency divergence of
$\Gamma_c^{\nu \nu'}$ occurs. This defines a curve $\widetilde{T}(U)$ with
a quite peculiar shape, where three
regions can be distinguished: (I) at very high $T$, the behavior is almost perfectly linear
$\widetilde{T} \propto \widetilde{U}$; (II) in the low T limit the curve
strongly bends, extrapolating for  $T \rightarrow 0$  at $\widetilde{U}(0)
\sim 1.5 \ll U_{\text{MIT}} \sim 3$; (III) at intermediate $T$  
the curve interpolates between these two regimes, with a
``re-entrance''  clearly affected by the presence of the MIT at larger
U (blue squares in Fig. \ref{fig:2}).
 We note that by increasing U much further than the $\widetilde{T}(U)$ curve,
one  eventually observes a divergence also of the local Bethe-Salpeter  
in the particle-particle channel (orange points in Fig. \ref{fig:2}),
while for all values of $T,U$ considered, no similar
divergence is found in the spin channel.

\noindent
{\sl Interpretation of the results.} In contrast to the case of the main diagonal structures of the
vertex functions, the interpretation of the low-frequency divergences of
$\Gamma_c^{\nu \nu'}$ is not directly related to the MIT. However,
even if at low $T$ the divergences take place in the metallic region of the
phase-diagram, the re-entrance shape of the $\widetilde{T}(U)$ curve is indeed
remarkably affected by the position of the MIT: The most natural
interpretation is, hence, that the shaded area in the phase
diagram defines the region where the precursor effects of the
MIT physics preclude the perturbative description and become a crucial
ingredient in determining the  properties of the system. This
interpretation is evidently supported by the fact that the signs of
the two-particle vertex functions are correctly predicted in perturbation
theory only up to the left-hand side of the $\widetilde{T}(U)$
curve. More generally, the $\widetilde{T}(U)$ curve can be identified as the limit of the region of applicability of schemes based on  the Baym-Kadanoff\cite{BKpr} functional $\Phi[G]$, since $\frac{\delta^2 \Phi}{\delta G^2} = \Gamma_c$ can no longer be defined on that line. At the same time, the low-frequency singularities of the vertex may render problematic the numerical evaluation of the Bethe-Salpeter equation to compute momentum-dependent DMFT response functions in specific regions of the phase-diagrams, suggesting the use of alternative procedures\cite{Nan2012}.

As the singularity of $\Gamma_{c}$ (and later on of $\Gamma_{pp}$) is not associated to simultaneous divergences in the other channels, the application of the local parquet equations\cite{bickers,Rohringer2012} allows to identify the ultimate root of these divergences in the {\sl fully} two-particle irreducible diagrams. Hence, this is an ``intrinsic'' divergence, deeply rooted in the diagrammatics and not generated by ladder scattering processes in {\sl any} channel. From a more physical point of view, the fact that the only irreducible vertex $\Gamma$ displaying no-singularities at low-frequencies is the spin one might also indicate the  emergent role played by preformed local magnetic moments as MIT precursors, even in regions where the metallic screening is rather effective.

We can go, however, also beyond these general considerations and
analyze the three regimes of the $\widetilde{T}(U)$ curve in detail,
also discussing the relation with the emergence of
some of the anomalous properties of the physics in the vicinity of the MIT.
The analysis of the high-T linear regime [(I) in Fig.\ \ref{fig:2}] of $\widetilde{T}(U)$ is
probably the easiest: Here $U, T \gg D$, and hence a connection
with the atomic limit ($D=0$) can be done: Using analytic
expressions\cite{Hafermann2009,Rohringer2012} for the reducible
two-particle vertex functions as an input
for Eq.\ (\ref{eq:Gamma}), we find that the low-frequency divergence of
$\Gamma_{c}^{\nu \nu'}$ occurs at $\widetilde{T}/\widetilde{U} = \frac{\sqrt{3} }{2 \pi}$ and that the eigenvector 
associated to the vanishing eigenvalue of $\chi_{c}^{\nu \nu'}$
has the particularly simple form: $\frac{1}{\sqrt{2}} (\delta_{\nu (\pi T)} -
\delta_{\nu (-\pi T)})$. As it is clear from the comparison with the red
dashed line in Fig. \ref{fig:2} this proportionality exactly matches
the high-$T$ linear behavior of our $\widetilde{T}(U)$ curve. 
Crossing this curve in its high-$T$ linear regime,
which extends indeed over a large portion
of the phase-diagram, corresponds to entering a region where the
thermal occupation of the high-energy doubly-occupied/empty states 
becomes negligible, letting the physics be dominated by the local
moments.  The connection with the local moment physics
 also holds for the low-$T$ region (II), though via a different mechanism:
 For $T \rightarrow 0$, the relevant energy scales are 
 the kinetic ($ \sim D$) and the potential ($U$) energy, whose competition
is regulated by quantum fluctuations. In this case, obviously, only numerical results are available:  We observe that the extrapolated
value of $\widetilde{U}(0) \sim 1.5$ falls in the same region (gray arrow in inset of Fig.\ \ref{fig:2}), where DMFT(NRG)\cite{NRG} see a first clear
separation of the Hubbard sub-bands from the central quasi-particle
peak in the spectral function $A(\omega)$.  We recall here that the formation of well-defined minima in $A(\omega)$ between the central quasi-particle peak and the Hubbard sub-bands, is directly connected with the anomalous phenomenon of the  appearance of kinks in the electronic self-energy and specific heat\cite{kinks2007}. At the same time, more recent DMFT(DMRG)\cite{Karski2008} data rather indicate that for $U \ge \widetilde{U}(0) \sim 1.5$ two sharp peak-features emerge at the inner edges of the Hubbard sub-bands, which, however, would be already visible at $U \ge 1$.       

Looking for a more analytical description of this scenario, we can consider the
DMFT solution of the much simpler Falicov-Kimball (FK) model\cite{FKREV}: Here one can exactly show that $\Gamma_{c}^{\nu\nu'}$ indeed diverges before the MIT is reached (precisely at: $\widetilde{U}^{\text{FK}} = \frac{1}{\sqrt{2}}
U_{\text{MIT}}^{\text{FK}}$).  However, for the FK  results, a direct relation with the formation of
the two minima in $A(\omega)$ cannot be completely identified, as 
the renormalization of the central peak is not captured in this scheme\cite{note_FK}.
Finally, an interesting observation can be made about  the most complicate
intermediate $T$ regime (III), where all energy scales
($D$, $U$, $T$) are competing:  Recent out-of-equilibrium calculations
for the Hubbard model have shown\cite{Eckstein2009}, that after a
quench of the interaction  (i.e., from $U=0$ to $U > 0$), the
system's relaxation occurs in two different (non-thermal) ways. The
changeover between these two regimes, however, appears for a given set of parameters $\bar{U}\sim 1.65$ and $T_{\text{eff}} \sim 0.4$,  in close
proximity of the vertex instability-line in our phase diagram. 
 
\noindent
{\sl Conclusions and outlook.} Our DMFT calculations have shown how the emergent (non-perturbative) precursor effects of the MIT determine a low-frequency divergence of the local Bethe-Salpeter equation in the charge channel. This allows for an {\sl unambiguous} identification of the regime,  where perturbation and Baym-Kadanoff functional theory break down, and where, at the same time, several anomalous properties are observed or predicted for correlated metals. Taking properly into account the physics emerging from the singularity of the two-particle vertex functions will represent one of the main challenges for future improvements of the theoretical many-body treatments at the precision level required by the increasingly higher experimental standards. 

\textit{Acknowledgments.} TS, GR and AT acknowledge financial support
from Austrian Science Fund (FWF) through the project I610-N16.
Numerical calculations have been performed on the Vienna
Scientific Cluster (VSC) and at the MPI-FKF in Stuttgart. We thank for insightful discussions:  M. Capone, K. Held, A. Georges, A. Katanin, C. Castellani,  M. Fabrizio, H. Hafermann, E. Gull, E. Kozik, J, Kune\^s, S. Andergassen, A. Valli, C. Taranto.


\begin{references}

\bibitem{MH}
 N. F. Mott, Rev.\ Mod.\ Phys.\ {\bf 40}, 677
(1968); {\sl Metal-Insulator Transitions} (Taylor \& Francis,
London, 1990); F. Gebhard, {\sl The Mott Metal-Insulator
Transition} (Springer, Berlin, 1997).

\bibitem{ImadaREV}  M. Imada {\sl et al.}, Rev. Mod. Phys. {\bf 70}, 1039 (1998).

\bibitem{DMFT} W.~Metzner and D.~Vollhardt, Phys. Rev. Lett. {\bf 62},  324  (1989); A. Georges and G. Kotliar, Phys. Rev. Lett. {\bf 45},  6479 (1992).   

\bibitem{DMFTREV}  A.~Georges {\sl et al.}, Rev. Mod. Phys. {\bf 68},  13 (1996). 

\bibitem{V2O3}  K. Held, {\sl et al.}, Phys. Rev. Lett. {\bf 86} 5345 (2001); 
A. I. Poteryaev, {\sl et al.}, Phys. Rev. B {\bf 76}, 085127 (2007).

\bibitem{Hubbard}
J.~Hubbard, \newblock Proc. Roy. Soc. London A {\bf 276}, 238 (1963).

\bibitem{LDADMFTREV} G. Kotliar {\sl et al.}, Rev. Mod. Phys. {\bf 78}, 865 (2006); K. Held, Adv. Phys. {\bf 56}, 829 (2007). 


\bibitem{kinks2007} K. Byczuk, {\sl et al.}, Nat. Phys. {\bf 3}, 168 (2007);
 A. Toschi {\sl et al.}, Phys. Rev. Lett. {\bf 102}, 076402 (2009).

\bibitem{Toschi2012} A. Toschi, {\sl et al.}, Phys. Rev. B {\bf 86},
  064411 (2012).

\bibitem{Eckstein2009} M. Eckstein, M. Kollar, and P. Werner
Phys. Rev. Lett. {\bf 103}, 056403 (2009); M. Schir\'{o} and M. Fabrizio,
Phys. Rev. B {\bf 83}, 165105 (2010).


\bibitem{energybal} C. Taranto {\sl et al.}, Phys. Rev. B {\bf 85}, 085124 (2012); A. Toschi, M. Capone, and C. Castellani, Phys. Rev. B {\bf 72}, 235118 (2005).

\bibitem{Raas2009} C. Raas and G. Uhrig, Phys. Rev. B, {\bf 79}, 115136 (2009).

\bibitem{Bullarev} R.~Bulla, \newblock Phys. Rev. Lett. {\bf 83} 136 (1999).

\bibitem{bluemer_thesis} N. Bl\"umer, Phd Thesis, (Augsburg, 2003).

\bibitem{Ciuchi2006} S. Ciuchi, G. Sangiovanni, and M. Capone,  Phys. Rev. B, {\bf 73}, 245114 (2006).

\bibitem{Dobro2011}  H. Terletska, {\sl et al.}, Phys. Rev. Lett. {\bf 107}, 026401  (2011).
 



\bibitem{Kunes2011} Jan Kune\^s, Phys. Rev. B {\bf 83}, 085102 (2011). 


\bibitem{Park2011} H. Park, K. Haule, and G. Kotliar,  Phys. Rev. Lett. {\bf 107}, 137007 (2011).

\bibitem{DGA} A. Toschi, A.A. Katanin, and K. Held,
 Phys. Rev. B {\bf 75}, 045118 (2007); G. Rohringer {\sl et al.}, Phys. Rev. Lett.  {\bf 107}, 256402 (2011).


\bibitem{DF}
 A. N. Rubtsov, M. I. Katsnelson, and A.I. Lichtenstein, Phys. Rev. B {\bf 77}, 033101 (2008);  H. Hafermann, {\sl et al.}, Phys. Rev. Lett. {\bf 102}, 206401 (2009).


\bibitem{note_D} As in DMFT the kinetic energy scale is controlled by $D$, DMFT results for different DOSes essentially coincides, provided the value of $D$ is the same. 


\bibitem{Rohringer2012} G. Rohringer, A. Valli, and A. Toschi, Phys. Rev. B, {\bf 86} 125114 (2012).


\bibitem{note_Tdep} In the (unpublished) appendix of arXiv:1104.3854v1, a two-particle vertex divergence was also reported for one temperature, whose position would have been controlled by $U$ rather than by $T$. This expectation is, however, not verified by our data of Fig. \ref{fig:2}. 


\bibitem{BKpr} G. Baym and L. P. Kadanoff, Phys. Rev. {\bf 124}, 287 (1961).


\bibitem{Nan2012}  Nan Lin, {\sl et. al}, Phys. Rev. Lett. {\bf 109}, 106401 (2012). 

\bibitem{bickers} D. Senechal {\sl et al.}, ''Theoretical Methods for Strongly Correlated Electrons'' (Chapter 6, N.E. Bickers, ''Self-Consistent Many-Body Theory for Condensed Matter Systems''), Springer (2003).


\bibitem{NRG} R.~ Zitzler, PhD Thesis (Augsburg, 2004); R. Bulla, Phys. Rev. Lett. {\bf 83} 136 (1999), we thank also R. Bulla for making the raw numerical renormalization group (NRG) data available to K. Held and his group.

\bibitem{Hafermann2009} H. Hafermann {\sl et. al}, Europhys. Lett. {\bf 85}, 27007 (2009).

\bibitem{Karski2008} M. Karski, C. Raas, and G. Uhrig, Phys. Rev. B, {\bf 77}, 075116 (2008). We thank P. Thunstr\" {o}m for making available his dynamical matrix renormalization group (DMRG) DMFT data.  

\bibitem{FKREV}  J. K. Freericks and V. Zlati\'c,  Rev. Mod. Phys. {\bf 75}, 1333  (2003).

\bibitem{note_FK} Note that the DMFT solution of the FK model,
  coinciding with the CPA, can only capture the MIT via a rigid separation
of the Hubbard band: Here the formation of a central minimum occurs
already at $U= \frac{1}{2} U_{MIT}^{FK}$.


\end{references}
\end{document}